\documentclass[12pt]{iopart}
\usepackage{amsfonts}

\begin{document}

\title{State transformations after quantum fuzzy measurements}
\author{Ioan Sturzu}

\begin{abstract}
Using a standard fuzzification procedure and the dynamical map in Heisenberg
picture, a new expression for the state transformation after a fuzzy filter
measurement, subject to covariance conditions, was obtained and some
calculations were done to distinguish its properties from the those of the
usual solution.
\end{abstract}

\maketitle

\address{"Transilvania" University, Department of Physics, Eroilor 29,
Brasov, Romania; email: sturzu@unitbv.ro}




\section{\label{intro}Introduction}

If a quantum system interaction with the surroundings is subject to some
restrictions like ''quasi-isolation'', i.e. it can interact up to a
non-relativistic conservative quasi-classical field of forces, its evolution
is given by the Schr\"{o}dinger equation, which yields the unitary dynamical
group (UG) $\{\widehat{U}_t\}_{t\in \Bbb{R}}$:
\[
|\Psi \left( t\right) \rangle =\widehat{U}_t|\Psi \left( 0\right) \rangle
\]
Otherwise, if it cannot be considered quasi-isolated, its evolution is
well-described by master-type equations (which are the most general
evolution equations which preserve the trace of the density operator $%
\widehat{\rho }$ \cite{lindblad}). They actually yield some completely
positive quantum dynamical semigroups (SG) \cite{davies,alicki} $\{\widehat{V%
}_t\}_{t\geq 0}$. However, if one can accept freely an ontological
interpretation of the conceptual apparatus of Quantum Physics in the first
case, one has to be careful when asserting ontological-type sentences for
the open quantum systems. This is due to pure quantum correlations which
appear between the system and the surroundings, which cannot be destructed
by any type of interaction. Of course, one can work perfectly, from the
epistemological point of vue \cite{schroek}, with the up-mentioned SG's and
master equations, but one has to be aware on the possible implications of
explicitly or hidden ontological presuppositions.

Due to Born probabilistic interpretation, one has to accept - in the same
epistemological perspective - the duality between states and observables,
which is clearly illustrated in the two pictures of quantum evolution
(Heisenberg and Schr\"{o}dinger) and by the concept of instrument in
Operational Quantum Physics (OQP) \cite{davies,schroek,helstrom}. In OPQ the
state transformer after a quantum measurement is defined as an instrument,
and this aims for the mathematical formalization of the so-called Projection
Postulate (a to much debated topic in Quantum Physics, often carrying an
ontological ballast). The evolution due to SG yield, in the Schr\"{o}dinger
picture, the mixing of initially pure states, while in Heisenberg picture a
sharp observable $\{\widehat{E}(B)\}_{B\in \mathcal{B}}$ goes to a fuzzy
observable $\{\widehat{F}(B)\}_{B\in \mathcal{B}}$ \cite{sturzu}. If one
consider the fuzzyfication map as a dynamical map in Heisenberg picture, one
can compute the corresponding dual dynamical map in Schr\"{o}dinger picture,
an interpret it as a state transformer - which is different to the state
transformer in OQP. This is a pure epistemological variant of OQP. Its study
is useful in a twofold perspective: its adequacy to the real experimental
physics could yield some changes in the present conceptual apparatus, while
its inadequacy could emphasize the ontological importance of the usual OQP.

\section{State transformations}

\subsection{General topics}

One of the main results of OQP is that one can define the state
transformation after a filter-measurement (state-preparation procedure)
using a quantum instrument \cite{davies}, that is a map: $\mathcal{E}%
(\bullet ,\bullet )$: $\mathcal{F\times B(H)\rightarrow B(H)}$, where ($%
\Omega ,\mathcal{F}$) is a Borel space on the set of experimental data $%
\Omega $, $\mathcal{B(H)}$ is the space of the bounded operators on the
Hilbert space $\mathcal{H}$, and for fixed $\hat{A}\in \mathcal{B(H)}$ $%
\mathcal{E}(\bullet ,\hat{A})$ is a $\sigma $-additive measure on $(\Omega ,%
\mathcal{F)}$, while the values $\mathcal{E}(B,\bullet ),$ $B\in \mathcal{F}$
are bounded completely positive operators $\mathcal{B(H)}\rightarrow
\mathcal{B(H)}$ for which:
\[
\mathcal{E}(\Omega ,\hat{I})=\hat{I}
\]
The evolution of the system is given, in Heisenberg image, by the map:
\[
\hat{A}\longmapsto \mathcal{E}(B,\hat{A})
\]
while the state transformation (Schr\"{o}dinger image) is given by the dual
instrument:
\begin{equation}
\widehat{\rho }\longmapsto \hat{\rho}_{post}(B;\hat{\rho})=\mathcal{E}^{*}(B,%
\hat{\rho})  \label{trsel}
\end{equation}
where $B\in \mathcal{B}(\Omega )$ is the Borel set of data, which the
filter-measurement is looking for. Beyond positivity and monotonicity, $%
\mathcal{E}^{*}(\bullet ,\widehat{\rho })$ is asked to fulfil the following
condition \cite{davies}:
\begin{equation}
Tr(\mathcal{E}^{*}(\Omega ,\widehat{\rho }))=Tr(\widehat{\rho })
\label{trpres}
\end{equation}
The sense of (\ref{trsel}) is given by the convention used in usual OQP
which accepts non-trace-preserving state transformers for $B\neq \Omega $ in
order to limit the mathematical apparatus only to linear transformations
\cite{bushlah}. The apparatus of OQP is trivial for sharp quantum
observables and is, in general, due to von Neumann \cite{vonNeumann}.
However, measurement is, by its very nature, a process of interaction which
cannot, in general, fulfil the conditions for the unitary evolution of the
investigated system, so one has to consider also, the case of fuzzy
observables \cite{bushlah}. The operators:
\[
\widehat{F}(B)=\mathcal{E}(B,\hat{1})
\]
are usually named \textit{effects} or \textit{fuzzy operators. }The
probability of obtaining the measurement result in the Borel set $B$ is
given in the terms of the instrument by:
\begin{eqnarray}
p(B) &=&Tr(\widehat{F}(B)\widehat{\rho })=Tr(\mathcal{E}(B,\hat{1})\widehat{%
\rho }))=  \label{instr} \\
&=&Tr(\mathcal{E}^{*}(B,\widehat{\rho }))=Tr(\hat{\rho}_{post}(B;\hat{\rho}))
\nonumber
\end{eqnarray}
In \cite{loubenets} and \cite{nielsenloubenets} it is given a representation
theorem for the most general form of a quantum instrument $\forall B\in
\mathcal{F},\forall \hat{A}\in \mathcal{B(H)}$:
\begin{equation}
\mathcal{E}(B,\hat{A})=\sum_{i=1}^{N_0}\alpha _i\int_{\omega \in
B}\sum_{n=1}^{N(\omega )}\hat{W}_{in}^{+}(\omega )\hat{A}\hat{W}_{in}(\omega
)\nu (d\omega ),  \label{instrmgen}
\end{equation}
where the operators $\{\hat{W}_{in}(\omega )\}_{i,n;\omega }$ have to
observe the relation:
\begin{equation}
\int_{\omega \in \Omega }\sum_{n=1}^{N(\omega )}\hat{W}_{jn}^{+}(\omega )%
\hat{W}_{in}(\omega )\nu (d\omega )=\delta _{ji}  \label{gennorm}
\end{equation}
(\ref{instrmgen}) and (\ref{gennorm}) are to be understood in the weak
sense. They give the most general form of the effects:
\begin{equation}
\widehat{F}(B)=\sum_{i=1}^{N_0}\alpha _i\int_{\omega \in
B}\sum_{n=1}^{N(\omega )}\hat{W}_{in}^{+}(\omega )\hat{W}_{in}(\omega )\nu
(d\omega )  \label{mgeneff}
\end{equation}
However, the set of these mathematical objects have not mandatory correspond
one-to-one to physical-significative objects, which have to observe some
restrictions (for example the case of regular effects, which are neither
less then $\frac 12\hat{1}$, nor greater than it).

The definition of the state transformer as an instrument aims to the
mathematical formalization of the so-called Projection Postulate (via
relation (\ref{instr})), which is a too much debated topic in Quantum
Mechanics literature. It is known \cite{ozawa}, that the conceptual status
of state projection (reduction), is still polemical in measurement theory
\cite{Joh93}, whereas the concept of (nonselective) state change is rather
straightforward, and can be easy handled in the Quantum Physics of Open
Systems \cite{davies,isar}.

In \cite{ozawa} it is shown that $\mathcal{E}(\Bbb{{Z},\widehat{\rho })}$
(nonselective state transformation) has a unique decomposition in (the
selective) state transformations:
\[
\mathcal{E}(\Bbb{{Z},\widehat{\rho })=}\sum\limits_{m\in \Bbb{Z}}\mathcal{E}%
(\{m\},\widehat{\rho }\Bbb{)}
\]
which it is claimed to support the (selective) state reduction without
assuming the Projection Postulate. However, the ideas presented above are
consistent, also, with a Quantum Mechanics without Projection Postulate
(without selection). If
\[
\widehat{E}(B)\mapsto \widehat{F}(B)
\]
is the dynamical map in Heisenberg picture (the sharp effects go to fuzzy
ones), one can write the probability using the Schr\H{o}dinger image of the
dynamical map (see eq. II.2.1 in \cite{alicki})

\begin{equation}
p(B)=Tr(\widehat{F}(B)\widehat{\rho })=Tr(\widehat{E}(B)\widehat{\rho }%
_{post}^{(E)})  \label{instrepstm}
\end{equation}
Of course, this is applicable only for those generalized effects which can
be obtained by fuzzyfication procedures from sharp ones. (But, for the
moment, there is no physical grounded theory for the most general form of
the physical effects, as it is done in \cite{loubenets} for the mathematical
objects (\ref{mgeneff}), while fuzzyfication of sharp obsevables is usual).

\subsection{Purely discrete case}

In \cite{davies,kraus} it is taken the case (\ref{instrmgen}) when $N_0=1$
and the measure $\nu $ is the Heaviside one :
\begin{equation}
\mathcal{E}(B,\hat{A})=\sum\limits_{m\in B}\widehat{\mathcal{A}}_m^{+}\hat{A}%
\widehat{\mathcal{A}}_m
\end{equation}
\begin{equation}
\mathcal{E}^{*}(B,\widehat{\rho })=\sum\limits_{m\in B}\widehat{\mathcal{A}}%
_m\widehat{\rho }\widehat{\mathcal{A}}_m^{+}  \label{instrgen}
\end{equation}
where:
\[
\sum\limits_{m\in \mathbb{Z}}\widehat{\mathcal{A}}_m^{+}\widehat{\mathcal{A}}%
_m=\widehat{1}
\]
On the next step, the case of a sharp observable with a purely discrete
spectrum $\{\widehat{E}_i\}_{i\in \Bbb{Z}}$ will be considered; it is
consistent with (\ref{instrgen}). The standard \textit{fuzzyfication}
procedure for a sharp observable is given by the relation \cite{zadeh,davies}%
:

\begin{equation}
\widehat{F}_\alpha =\sum\limits_{m\in \mathbb{Z}}{\omega _{\alpha m}}%
\widehat{E}_m\quad \forall \alpha \in \Bbb{Z}  \label{eff}
\end{equation}
\begin{equation}
\widehat{F}(B)=\sum\limits_{\alpha \in B}\sum\limits_{m\in \mathbb{Z}}{%
\omega _{\alpha m}}\widehat{E}_m  \label{effB}
\end{equation}
where the positive constants $\left\{ \omega _{\alpha m}\right\} _{\alpha
,m\in \Bbb{Z}}$ have to observe the relation:
\begin{equation}
\sum\limits_{\alpha \in \mathbb{Z}}\omega _{\alpha m}=1\quad \forall m\in
\Bbb{Z}  \label{norm}
\end{equation}
(simple stochasticity) and may depend on some \textit{fuzzyfication
parameters} (like $\sigma $ which, for the moment, will be dropped).From (%
\ref{instr}), (\ref{eff}) and (\ref{instrgen}) one gets:
\[
\widehat{\mathcal{A}}_m\widehat{\mathcal{A}}_m^{+}=\sum\limits_{k\in %
\mathbb{Z}}{\omega _{mk}}\widehat{E}_k=\widehat{F}_m
\]
Here, usually, one takes a non-complex solution:
\begin{equation}
\widehat{\mathcal{A}}_m=\widehat{\mathcal{A}}_m^{+}=\sum\limits_{k\in %
\mathbb{Z}}\sqrt{{\omega _{mk}}}\widehat{E}_k=\sqrt{\widehat{F}_m}
\label{aemde}
\end{equation}
so (\ref{instrgen}) becomes:
\begin{equation}
\mathcal{E}^{*}(B,\widehat{\rho })=\sum\limits_{m\in B}\sum\limits_{k,n\in %
\mathbb{Z}}\sqrt{{\omega _{mk}\omega _{mn}}}\widehat{E}_k\widehat{\rho }%
\widehat{E}_n
\end{equation}
\begin{equation}
\mathcal{E}^{*}(B,\widehat{\rho })=\sum\limits_{m\in B}\sqrt{\widehat{F}_m}%
\widehat{\rho }\sqrt{\widehat{F}_m}  \label{mainobj}
\end{equation}
The result (\ref{mainobj}) is often taken as the main objective of OQP (see
page 138 of \cite{schroek}). Of course, in the sharp situation, when ${%
\omega _{mk}=\delta _{mk}}$, one gets the usual von Neumann result:
\begin{equation}
\mathcal{E}_{vN}^{*}(B,\widehat{\rho })=\sum\limits_{m\in B}\widehat{E}_m%
\widehat{\rho }\widehat{E}_m  \label{vonNeum}
\end{equation}

Meanwhile, using (\ref{instrepstm}) the probabilities of measuring the fuzzy
observable (\ref{eff}) $\{\widehat{F}_\alpha \}_{\alpha \in \Bbb{Z}}$ for a
system in the state $\widehat{\rho }$ are given by:
\begin{equation}
p_\alpha =Tr(\widehat{F}_\alpha \widehat{\rho })=\sum\limits_{m\in \mathbb{Z}%
}\omega _{\alpha m}Tr(\widehat{E}_m\widehat{\rho })=\sum\limits_{m\in %
\mathbb{Z}}\omega _{\alpha m}\rho _{mm}  \label{prob}
\end{equation}
and the dynamical map in Schr\H{o}dinger image can be to write using (\ref
{prob}) in the terms of a completely positive map, similar to that of the
dual instrument (\ref{instrgen}):
\[
\widehat{\rho }\mapsto \widehat{\rho }_{post}^{(E)}=\mathcal{E}(\Bbb{{Z},%
\widehat{\rho })=}\sum\limits_{m\in \mathbb{Z}}\widehat{\mathcal{A}}_m^{(E)}%
\widehat{\rho }\widehat{\mathcal{A}}_m^{(E)+}
\]
\[
p_\alpha =Tr(\widehat{E}_\alpha \widehat{\rho }_{post}^{(E)})=Tr(|\alpha
\rangle \langle \alpha |\sum\limits_{m\in \mathbb{Z}}\widehat{\mathcal{A}}%
_m^{(E)}\widehat{\rho }\widehat{\mathcal{A}}_m^{(E)+})\quad \forall \alpha
\in \Bbb{Z}
\]
\begin{equation}
\sum\limits_{m\in \mathbb{Z}}\omega _{\alpha m}\rho _{mm}=\sum\limits_{m\in %
\mathbb{Z}}\langle \alpha |\widehat{\mathcal{A}}_m^{(E)}\widehat{\rho }%
\widehat{\mathcal{A}}_m^{(E)+}|\alpha \rangle \quad \forall \alpha \in \Bbb{Z%
}  \label{eqdef}
\end{equation}
In (\ref{eqdef}) one can identify:
\begin{eqnarray}
\widehat{\mathcal{A}}_m^{(E)} &=&\sum\limits_{k\in \mathbb{Z}}\sqrt{{\omega
_{km}}}|k\rangle \langle m|  \label{aeme} \\
\widehat{\mathcal{A}}_m^{(E)+} &=&\sum\limits_{k\in \mathbb{Z}}\sqrt{{\omega
_{km}}}|m\rangle \langle k|  \nonumber
\end{eqnarray}
which are clearly different from (\ref{aemde}).

One is looking for a positive operator $\widehat{\mathcal{F}}_d$ which
yields:
\begin{equation}
\omega _{km}=\langle m|\widehat{\mathcal{F}}_d^{+}|k\rangle \langle k|%
\widehat{\mathcal{F}}_d|m\rangle   \label{efde}
\end{equation}
If one takes the real-positive solution $\langle k|\widehat{\mathcal{F}}%
_d|m\rangle =\sqrt{\omega _{km}}$, one can write:
\begin{equation}
\widehat{\rho }_{post}^{(E)}=\widehat{\mathcal{F}}_d\sum\limits_{m\in %
\mathbb{Z}}\widehat{E}_m\widehat{\rho }\widehat{E}_m\widehat{\mathcal{F}}%
_d^{+}  \label{instris}
\end{equation}
which can be written as (\ref{instrgen}) for
\begin{eqnarray}
\widehat{\mathcal{A}}_m^{(E)} &=&\widehat{\mathcal{F}}_d\widehat{E}_m \\
\widehat{\mathcal{A}}_m^{(E)+} &=&\widehat{E}_m\widehat{\mathcal{F}}_d^{+}
\nonumber
\end{eqnarray}
The ideal case of von Neumann can be obtained, again, for $\widehat{\mathcal{%
F}}_d=\widehat{1}$.

Using (\ref{efde}) one can write (\ref{aemde}) as:
\[
\widehat{\mathcal{A}}_m=\widehat{\mathcal{A}}_m^{+}=\sum\limits_{k\in %
\mathbb{Z}}\langle m|\widehat{\mathcal{F}}_d|k\rangle \widehat{E}_k
\]

\subsubsection{Example\label{poisson}}

An interesting case is that of a (symmetrical) Gaussian convolution in (\ref
{eff}):
\begin{equation}
{\omega _{km}=}\frac 1{\Psi _0(\sigma )}\exp (-\frac{(k-m)^2}{\sigma ^2})
\label{gaussd}
\end{equation}
where $\Psi _0(\sigma )=\sum\limits_{m\in \mathbb{Z}}\exp (-\frac{k^2}{%
\sigma ^2})$ is a function whose properties can be found using Poisson
summation formula \cite{lighthill,is} ($\Psi _0(\sigma )\cong 1$ for $\sigma
\in [0;0.4]$ and $\Psi _0(\sigma )\cong \sqrt{\pi }\sigma $ for $\sigma \in
[0.8;\infty )$. Also, one can define $\Psi _{1/2}(\sigma )=\sum\limits_{m\in %
\mathbb{Z}}\exp (-\frac{(k+\frac 12)^2}{\sigma ^2})$ and find that $\Psi
_{1/2}(\sigma )\cong 0$ for $\sigma \in [0;0.2]$ and $\Psi _{1/2}(\sigma
)\cong \sqrt{\pi }\sigma $ for $\sigma \in [0.8;\infty )$). In this case one
has:
\begin{equation}
\langle k|\widehat{\mathcal{F}}_d|m\rangle =\frac 1{\sqrt{\Psi _0(\sigma )}%
}\exp (-\frac{(k-m)^2}{2\sigma ^2})  \label{gaussd1}
\end{equation}

\subsection{Purely continuous case}

For the purely continuous spectrum, say the position operator case:
\[
\widehat{x}=\int\limits_{\Bbb{R}}x\cdot \widehat{E}(dx)=\int\limits_{\Bbb{R}%
}x\cdot |x\rangle \langle x|dx
\]
one takes in (\ref{instrmgen}) $\Omega =\Bbb{R}$ and $N(\omega )=N_0=1$. For
the fuzzy effect one takes:
\[
\widehat{F}(B)=\int\limits_{\Bbb{R}}(f\circ \chi _B)(x)\widehat{E}%
(dx)=\int\limits_Bd\lambda \int\limits_{\Bbb{R}}f(\lambda -x)\widehat{E}(dx)
\]
where $\int\limits_{\Bbb{R}}f(x)dx=1$ is a condition similar to (\ref{norm})
and $f(x)>0\quad \forall x\in \Bbb{R}$. The next step is looking for the
positive operator $\widehat{\mathcal{F}}_c$ for which:
\[
f(\lambda -x)=\langle x|\widehat{\mathcal{F}}_c^{+}|\lambda \rangle \langle
\lambda |\widehat{\mathcal{F}}_c|x\rangle
\]
Of course:
\begin{equation}
\langle \lambda |\widehat{\mathcal{F}}_c|x\rangle =\sqrt{f(\lambda -x)}
\label{elmatr}
\end{equation}
and looking for a state transformer like (\ref{instrepstm}) one finds:
\begin{equation}
\widehat{\rho }\mapsto \widehat{\rho }_{post}^{(E,c)}=\int\limits_{\Bbb{R}}dx%
\widehat{\mathcal{F}}_c|x\rangle \langle x|\widehat{\rho }|x\rangle \langle
x|\widehat{\mathcal{F}}_c^{+}  \label{instrcont}
\end{equation}
The OPQ state transformer is (equations 4.6.3 and 4.6.4 in \cite{davies}):
\begin{equation}
\widehat{\rho }_{post}^{(c)}=\int\limits_{\Bbb{R}}dx\widehat{\mathcal{A}}_x%
\widehat{\rho }\widehat{\mathcal{A}}_x^{+}  \label{sttrdav}
\end{equation}
where:
\[
\langle \lambda |\widehat{\mathcal{A}}_x|\Psi \rangle =\sqrt{f(\lambda -x)}%
\Psi (\lambda )\quad \forall \lambda \in \Bbb{{R},\forall }\Psi \Bbb{\in
\mathcal{H}}
\]
\[
\widehat{\mathcal{A}}_x=\widehat{\mathcal{A}}_x^{+}=\int\limits_{\Bbb{R}}%
\sqrt{f(x-y)}\widehat{E}(dy)=\int\limits_{\Bbb{R}}\langle x|\widehat{%
\mathcal{F}}_c|y\rangle \widehat{E}(dy)
\]
while (\ref{instrcont}) can be written in the form:
\begin{equation}
\widehat{\rho }_{post}^{(E,c)}=\int\limits_{\Bbb{R}}dx\widehat{\mathcal{A}}%
_x^{(E)}\widehat{\rho }\widehat{\mathcal{A}}_x^{(E)+}  \label{instrcis}
\end{equation}
where:
\begin{eqnarray*}
\widehat{\mathcal{A}}_x^{(E)} &=&\widehat{\mathcal{F}}_c|x\rangle \langle x|
\\
\widehat{\mathcal{A}}_x^{(E)+} &=&|x\rangle \langle x|\widehat{\mathcal{F}}%
_c^{+}
\end{eqnarray*}

\subsubsection{Example}

As an example it will be discussed only the continuous analogue of (\ref
{gaussd1}) which is given by:
\begin{equation}
\langle x|\widehat{\mathcal{F}}_c|x^{\prime }\rangle =\frac 1{\sqrt{\sqrt{%
\pi }\sigma }}\exp (-\frac{(x-x^{\prime })^2}{2\sigma ^2})
\end{equation}
\[
\widehat{\mathcal{F}}_c=\sqrt{2\sqrt{\pi }\sigma }\exp (-\frac 12\sigma ^2%
\widehat{k}^2)
\]
where $\widehat{k}$ is the wave-number operator, conjugate to $\widehat{x}$%
.\medskip

In \cite{davies} it is shown that $\mathcal{E}^{*}(B,\widehat{\rho })$, the
corresponding instrument to (\ref{sttrdav}), is a solution of the covariance
condition:
\[
\mathcal{E}^{*}(B+a,\widehat{\rho })=\widehat{U}_a^{+}\mathcal{E}^{*}(B,%
\widehat{U}_a\widehat{\rho }\widehat{U}_a^{+})\widehat{U}_a
\]
where $\widehat{U}_a$ is the shift-operator on $\Bbb{R}$,
\[
\widehat{U}_a=\exp (ia\widehat{k})
\]

Starting from (\ref{elmatr}) one can calculate the matrix elements in the
wave-numbers space, and find that:
\[
\langle k|\widehat{\mathcal{F}}_c|k^{\prime }\rangle =(\sqrt{f})^{\symbol{126%
}}\delta (k-k^{\prime })
\]
\[
\widehat{\mathcal{F}}_c=(\sqrt{f})^{\symbol{126}}(\widehat{k})
\]
where $(\sqrt{f})^{\symbol{126}}$ is the Fourier Transform of $\sqrt{f(x)}$.
One follows that $[\widehat{\mathcal{F}}_c,\widehat{U}_a]=0$, so one can
easy prove that the completely positive map (\ref{instrcis}) is also
covariant w.r.t. space translations.

\section{Calculations with the two variants of state-transformation}

\subsection{Discrete case}

Calculating the moments of the observable $\widehat{E}$ for the two states (%
\ref{sttrdav}) and (\ref{instris}) one obtains:
\begin{eqnarray*}
M_1^{(O)} &=&Tr(\widehat{\rho }_{post}^{(c)}\sum\limits_{m\in \mathbb{Z}%
}m|m\rangle \langle m|)=\sum\limits_{m,k\in \mathbb{Z}}m{\omega _{km}}\rho
_{mm}= \\
&=&\sum\limits_{k\in \mathbb{Z}}{\omega _{km}}\sum\limits_{m\in \mathbb{Z}%
}m\rho _{mm}=\sum\limits_{m\in \mathbb{Z}}m\rho _{mm}= \\
&=&Tr(\widehat{\rho }\sum\limits_{m\in \mathbb{Z}}m|m\rangle \langle m|)=M_1
\end{eqnarray*}
One can notice that $M_1^{(O)}$, and also all the superior moments, do not
depend on the fuzzyfication parameters. In the other case, by the contrary,
one has:
\begin{eqnarray*}
M_1^{(E)} &=&Tr(\widehat{\rho }_{post}^{(E,c)}\sum\limits_{m\in \mathbb{Z}%
}m|m\rangle \langle m|)=\sum\limits_{m,k\in \mathbb{Z}}m{\omega _{mk}}\rho
_{kk}= \\
&=&\sum\limits_{k\in \mathbb{Z}}k\rho _{kk}+\sum\limits_{m,u\in \mathbb{Z}}u{%
\omega _{m+u,m}}\rho _{mm}
\end{eqnarray*}
Let $\sum\limits_{u\in \mathbb{Z}}u{\omega _{m+u,m}=}M_1({\omega ,m)}$. One
will consider only homogeneous distributions, that is ${\omega _{m+u,m}=}%
\omega _u$, for which $M_1({\omega ,m)=}M_1^{({\omega )}}$. Then:
\[
M_1^{(E)}=M_1+M_1^{({\omega )}}
\]
and, in general:

\[
M_n^{(E)}=\sum\limits_{k=0}^n\left(
\begin{array}{l}
k \\
m
\end{array}
\right) M_kM_{n-k}^{({\omega )}}
\]
which is a well-known formula of statistical physics.

Some straightforward calculations can be done for the linear entropy \cite
{isars} (see also papers: \cite{fizrev,fizlet}):
\begin{equation}
S=Tr(\widehat{\rho }-\widehat{\rho }^2)=1-Tr(\widehat{\rho }^2)  \label{entr}
\end{equation}

For (\ref{sttrdav}) one has:
\begin{equation}
S^{(O)}=1-\sum\limits_{n,k\in \mathbb{Z}}(\sum\limits_{m\in \mathbb{Z}}\sqrt{%
{\omega _{mk}\omega _{mn}}})^2\rho _{kn}\rho _{nk}  \label{entrdavd}
\end{equation}
while for (\ref{instris}):
\begin{equation}
S^{(E)}=1-\sum\limits_{n,k\in \mathbb{Z}}(\sum\limits_{m\in \mathbb{Z}}\sqrt{%
{\omega _{mk}\omega _{mn}}})^2\rho _{nn}\rho _{kk}  \label{entrepd}
\end{equation}

\subsubsection{Example}

For (\ref{gaussd}) one has: $M_1^{(E)}=M_1$, but for $n>1$, $M_n^{(E)}\neq
M_n$. The entropies (\ref{entrdavd}) and (\ref{entrepd}) become:
\[
S^{(O)}=1-\sum\limits_{k\in \mathbb{Z}}\sum\limits_{m\in \mathbb{Z}}[\rho
_{k,k+2m}\cdot \rho _{k+2m,k}\cdot \exp (-2\frac{m^2}{\sigma ^2})+
\]
\[
+(\frac{\Psi _{1/2}(\sigma )}{\Psi _0(\sigma )})^2\rho _{k,k+2m+1}\cdot \rho
_{k+2m+1,k}\cdot \exp (-2\frac{(m+\frac 12)^2}{\sigma ^2})]
\]
\[
S^{(E)}=1-\sum\limits_{k\in \mathbb{Z}}\sum\limits_{m\in \mathbb{Z}}[\rho
_{k,k}\cdot \rho _{k+2m,k+2m}\cdot \exp (-2\frac{m^2}{\sigma ^2})+
\]
\[
+(\frac{\Psi _{1/2}(\sigma )}{\Psi _0(\sigma )})^2\rho _{k,k}\cdot \rho
_{k+2m+1,k+2m+1}\cdot \exp (-2\frac{(m+\frac 12)^2}{\sigma ^2})]
\]
For a fuzzy initial state like:
\begin{equation}
\widehat{\rho }=\widehat{\mathcal{F}}_d^{(\alpha )}|a\rangle \langle a|%
\widehat{\mathcal{F}}_d^{(\alpha )+}\quad a\in \mathbb{Z},\alpha \in %
\mathbb{R}_{+}  \label{gausinstat}
\end{equation}
(\ref{entrdavd}) and (\ref{entrepd}) are equal even in the general case:
\[
S^{(O)}=S^{(E)}=1-\sum\limits_{n,k\in \mathbb{Z}}(\sum\limits_{m\in %
\mathbb{Z}}\sqrt{{\omega _{mk}^{(\sigma )}\omega _{mn}^{(\sigma )}}})^2{%
\omega _{n,a}^{(\alpha )}\omega _{k,a}^{(\alpha )}}
\]
which for (\ref{gaussd}) is:
\begin{eqnarray*}
&&1-\frac 1{\left( \Psi _0(\alpha )\right) ^2}[\Psi _0(\frac \alpha 2)\Psi
_0(\frac{\alpha \cdot \sigma }{\sqrt{2\cdot (\alpha ^2+2\sigma ^2)}})+ \\
&&+(\frac{\Psi _{1/2}(\sigma )}{\Psi _0(\sigma )})^2\Psi _{1/2}(\frac \alpha
2)\Psi _{1/2}(\frac{\alpha \cdot \sigma }{\sqrt{2\cdot (\alpha ^2+2\sigma ^2)%
}})]
\end{eqnarray*}
Using the approximations from the example (\ref{poisson}), for a sharp
initial position $(\alpha <0.2)$, the result is independent on the
fuzzyfication parameter $\sigma $, $S=0$ (the state remains almost pure),
while for $\alpha >1.6$ (fuzzy initial position), the result is $1-\frac
\alpha {\sqrt{\pi }}$ for sharp measurements $(\sigma <0.2)$ and $1-\frac
\sigma {\sqrt{2\cdot (\alpha ^2+2\sigma ^2)}}$ for unsharp measurements (if
still $\alpha \ll \sigma $ this becomes the maximum value for (\ref{entr}), $%
S=\frac 12$)

\subsection{Continuous case}

One will obtain similar formulas for the position operator $\widehat{x}%
=\int\limits_{\Bbb{R}}x\cdot \widehat{E}_{(x)}(dx).$ For the canonically
conjugate operator: $\widehat{k}=\int\limits_{\Bbb{R}}k\cdot \widehat{E}%
_{(k)}(dk)$ one has:
\[
M_1^{(O)}(\widehat{E}_{(k)})=Tr(\widehat{\rho }_{post}^{(c)}\int\limits_{%
\Bbb{R}}k\cdot \widehat{E}_{(k)}(dk))=
\]
\begin{eqnarray}
&=&Tr(\int\limits_{\Bbb{R}}dx\widehat{\mathcal{A}}_x\widehat{\rho }\widehat{%
\mathcal{A}}_x^{+}\int\limits_{\Bbb{R}}k\cdot \widehat{E}_{(k)}(dk))=
\label{momko} \\
&=&i\int\limits_{\Bbb{R}}dy[(\frac{\partial \rho }{\partial y^{\prime }}%
)_{(y,y^{\prime }=y)}-\frac 12\rho (y,y)\int\limits_{\Bbb{R}}dx(\frac{%
\partial f}{\partial x^{\prime }})_{(x^{\prime }=x-y)}]  \nonumber
\end{eqnarray}
\begin{eqnarray}
M_1^{(E)}(\widehat{E}_{(k)}) &=&Tr(\widehat{\rho }_{post}^{(E,c)}\int%
\limits_{\Bbb{R}}k\cdot \widehat{E}_{(k)}(dk))=  \nonumber \\
&=&Tr(\int\limits_{\Bbb{R}}dx\widehat{\mathcal{A}}_x^{(E)}\widehat{\rho }%
\widehat{\mathcal{A}}_x^{(E)}\int\limits_{\Bbb{R}}k\cdot \widehat{E}%
_{(k)}(dk))=  \nonumber \\
&=&-\frac i2\int\limits_{\Bbb{R}}dy\cdot \rho (y,y)\int\limits_{\Bbb{R}}dx(%
\frac{\partial f}{\partial x^{\prime }})_{(x^{\prime }=x-y)}  \label{momke}
\end{eqnarray}
The difference between (\ref{momko}) and (\ref{momke}) is given by the term $%
i\int\limits_{\Bbb{R}}(\frac{\partial \rho }{\partial y^{\prime }}%
)_{(y,y^{\prime }=y)}dy$, which is a memory term for the quantum coherence
of the initial state (prior to the fuzzy measurement of position), which,
however, does not depend on the fuzzyfication parameters, as one may expect.

\section{Conclusions}

Starting with a fuzzyfication process as a dynamical map in Heisenberg
picture, an alternate expression for the state transformer after a
(fuzzy-)filter measurement was proposed. It is a completely positive map,
similar to the dual quantum instrument of OQP, which is, also, subject to
covariance conditions. Calculations distinguished some different predictions
to that of the OPQ solution.

\ack

The author is grateful to Dr. H. Scutaru for stimulating discussions.


\section*{References}

\begin{thebibliography}{99}
\bibitem{lindblad}  Lindblad G 1976 \textit{Comm. Math. Phys.} \textbf{48} p
199

\bibitem{davies}  Davies E B 1976 \textit{Quantum Theory of Open Systems}
(London: Academic Press)

\bibitem{alicki}  Alicki R and Lendi K 1987 \textit{Quantum Dynamical
Semigroups and Applications} (Berlin: Springer-Verlag)

\bibitem{schroek}  Schroek F 1996 \textit{Quantum Mechanics on Phase Space}
(Dordrecht: Kluwer)

\bibitem{helstrom}  Helstrom C W 1976 \textit{Quantum Detection and
Estimation Theory} (New York: Academic Press)

\bibitem{sturzu}  Sturzu I 2002a Topics on the stochastic treatment of the
evolution of an open quantum system \textit{Preprint:} quant-ph/0204014 (to
appear in \textit{Romanian Journal of Physics})

\bibitem{loubenets}  Loubenets E R 2002 \textit{J.\ Phys.\ A} \textbf{34} p
7639-7675

\bibitem{nielsenloubenets}  Barndorff-Nielsen O E and Loubenets E R 2002
\textit{J.\ Phys.\ A} \textbf{\ 35 }p\textbf{\ }565-588

\bibitem{bushlah}  Busch P Grabowski M and Lahti P 1995 \textit{Operational
Quantum Physics} LNP m31, (Berlin: Springer-Verlag)

\bibitem{vonNeumann}  von Neumann J 1955 \textit{Mathematical foundations of
quantum mechanics} (Princeton: Princeton University Press)

\bibitem{zadeh}  Zadeh L A 1965 \textit{Information and Control}\emph{\/}
\textbf{8} p 338-353

\bibitem{kraus}  Kraus K 1997 Annals Phys. \textbf{64} p 311

\bibitem{ozawa}  Ozawa M 1997 Annals Phys. \textbf{259} p 121

\bibitem{Joh93}  Johnston J~R 1993 \textit{Phys.\ Rev.\ A} \textbf{48} p 2497

\bibitem{lighthill}  Lighthill M J 1958 \textit{Introduction to Fourier
Analysis and Generalised Functions} (Cambridge: Cambridge University Press)

\bibitem{is}  Sturzu I 2002b On the function... \textit{Preprint:}
quant-ph/0203006 (to appear in \textit{Bulletin of the Transilvania
University of Brasov})

\bibitem{isar}  Isar A, Sandulescu A, Scutaru H, Stefanescu E and Scheid W
1994 \textit{Int. J. Mod. Phys.} E \textbf{3} p 635-714

\bibitem{isars}  Isar A 1999 \textit{Fortschritte der Physik} \textbf{47} p
855-879

\bibitem{fizrev}  Paraoanu Gh S and Scutaru H 1998 \textit{Phys. Rev.} A
\textbf{58} p 869

\bibitem{fizlet}  Paraoanu Gh S and Scutaru H 1998 \textit{Phys. Lett.} A
\textbf{238} p 219
\end{thebibliography}
\end{document}